\documentclass[b5paper,twoside]{jpconf}
\usepackage{graphicx}
 
\usepackage{color}
\usepackage{hyperref}
\hypersetup{
            colorlinks=true,
            linkcolor=black,
            filecolor=black,
            citecolor=black,
            urlcolor=blue
           }
\definecolor{Dred}{rgb}{0.312,0.070,0.070}
\definecolor{Dblue}{rgb}{0.070,0.070,0.312}
\definecolor{Dgreen}{rgb}{0.070,0.312,0.070}
\definecolor{Db}{rgb}    {0.050,0.0,0.320}

\newcommand{\Blb}[1]{\textcolor{Dblue}{\bf #1}}

\newcommand{\web}[1]{\Blb{\url{#1}}}

\begin{document}
\title{VLBA observations of a complete sample of 2MASS galaxies} 

\author[Condon et al.]{\hspace{-5em}Jim Condon$^1$, Jeremy Darling$^2$, 
                        Yuri Y.~Kovalev$^3$, Leonid Petrov$^4$}

\address{$^1$National Radio Astronomy Observatory, Charlottesville, VA, USA}
\address{$^2$University of Colorado, USA}
\address{$^3$Astro Space Center, Lebedev Physical Institute, Russia}
\address{$^4$Astrogeo Center, USA}

\ead{jcondon@nrao.edu}

\begin{abstract}
    We are using the VLBA at 8~GHz to observe a sample of 834 nearby
    2MASS galaxies that are stronger than 100~mJy in the NVSS. The
    goals of the project are to detect (1) supermassive black holes
    significantly offset from the IR positions of the host galaxy
    bulges and (2) binary black holes.
\end{abstract}
\par\vspace{-6ex}\par

\section{Problem statement}

  Nearly all galaxy bulges contain supermassive black holes (SMBHs), 
so galaxy assembly by hierarchical merging should produce wide pairs 
of slowly inspiraling SMBHs that either ``stall'' as tight binaries with  
1~pc separation or merge and are kicked away from the galaxy nucleus 
by anisotropic gravitational radiation. We have begun a systematic VLBA 
search for off-nuclear inspiraling or recoiling SMBHs and for tight SMBH 
pairs in a complete sample of 923 nearby (D  $\sim 200$ Mpc) 2MASS galaxies 
containing NVSS radio sources stronger than 100~mJy. This survey 
simultaneously addresses three scientific problems: (1)~SMBH/galaxy 
co-evolution implied by the SMBH/bulge mass correlation, (2)~the 
``merger tree'' theory of SMBH evolution, and (3)~the expected contribution 
of merging SMBH binaries to the gravitational-wave background sought by
LISA and the NANOGrav pulsar timing experiment. 

  We search for offset or multiple SMBHs in a large complete sample 
by observing radio sources with the VLBA at 8~GHz (X band) to (1)~filter 
out extended emission from kiloparsec-scale radio jets and compact 
starbursts whose brightness temperatures never exceed $T \sim~10^5$~K 
\cite{r:con92}; (2) resolve stalled SMBH binaries; and (3)~favor compact 
flat-spectrum cores located very close to the SMBHs over extended 
steep-spectrum emission. Our complete sample of nearby stellar bulges 
begins with all 2MASS galaxies brighter than $K20fe = 12.25$ at 
$\lambda = 2.2 \mu$m, the wavelength at which luminosity is a good tracer 
of total stellar mass. The typical distance to these galaxies is 200 Mpc. 

  For our SMBH tracer, we use the compact cores of the 923 NVSS sources 
stronger than 100~mJy at 1.4~GHz identified with these 2MASS galaxies. 
Among them, 89 were observed in prior experiments. The luminosity of a 
100~mJy source at $D \sim\! 200$~Mpc is only 
$L \sim\! 10^{24}$~W Hz${}^{-1}$, but  90\% of these galaxies are 
radio-loud compared with the FIR/radio correlation of star-forming 
galaxies, so they almost certainly contain AGNs. 

\section{Observations}

  Each target source is observed in two scans at 8 intermediate
  frequencies spread over 7.90--8.89~GHz. Observations are made with
  phase calibrators in the C--T--C mode with $2 \times 60$~s on a
  calibrator and 320~s on a target in each scan. Observing sessions
  are scheduled by the VLBA array operator using an automatic
  algorithm at the lowest priority during periods of time when no
  other experiments can run due to various constraints.  Observations
  of individual calibrator/target pairs are scheduled in the so-called
  absolute astrometry mode with a set of four different atmosphere
  calibrator sources observed every 1.5~hours. This hybrid observing
  mode allows us to process the data both as absolute astrometry
  experiments using wide-band group delays determined with the
  wide-band baseline fringe fitting algorithm~\cite{r:vgaps} and as
  differential astrometry experiments.

  By October 2011, 328~hours of observing time were allotted. In total, 
733  of the 834 sources were observed, and among them 401 were 
detected in the baseline mode. The baseline detection limit is 6--9~mJy, 
depending on the weather.
  
\begin{figure}[h]
  \includegraphics[width=0.61\textwidth]{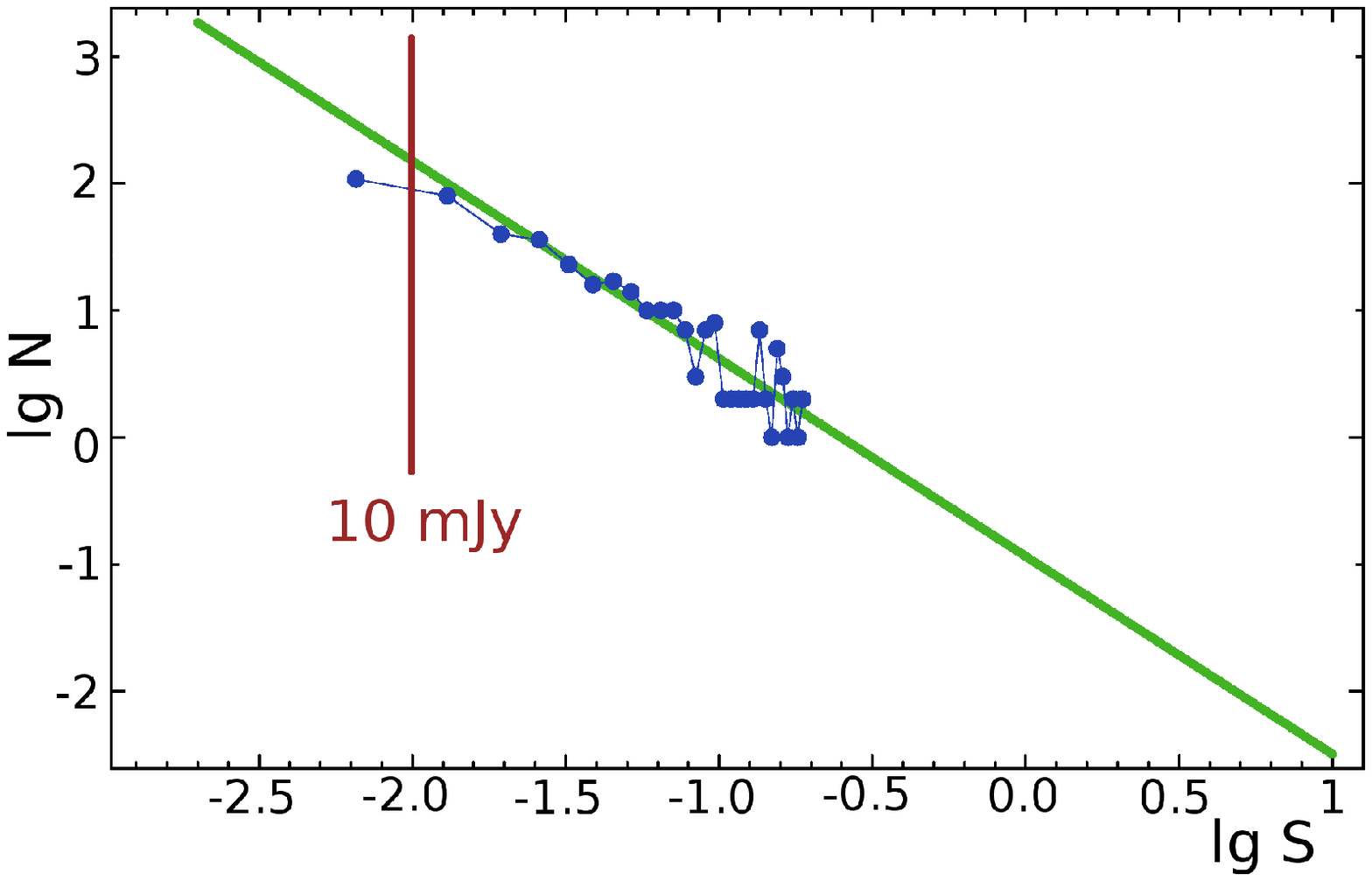}
  \hspace{0.005\textwidth}
  \raisebox{1ex}{\includegraphics[width=0.38\textwidth]{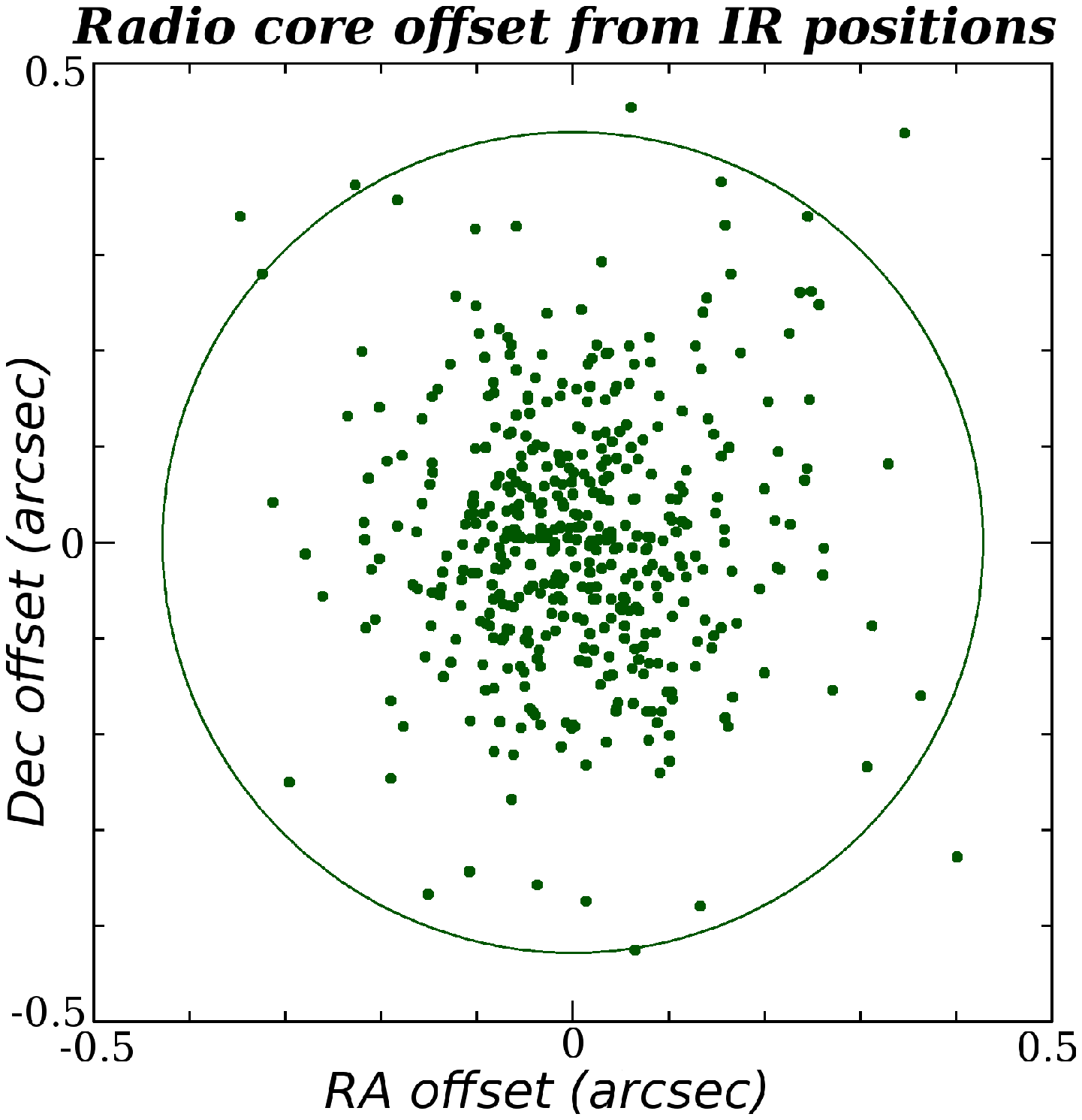}}
  \par\vspace{-2ex}\par
  \caption{\label{f:flux} {\it Left:} the distribution of correlated
    flux densities of unresolved radio components in our sample of
    galaxies. The units for flux densities are Jansky. The straight fitting 
      line has slope $-1.55$. {\it Right:} the offsets of the radio cores 
      from the IR positions. The circle with radius $0''.43$ corresponds to 
      the 99.9\% confidence level of a source offset due to random errors 
      of the 2MASS catalogue.}
\end{figure}

  The distribution of the correlated flux densities is shown in 
Figure~\ref{f:flux}.

  The median accuracy of VLBI positions is 0.6 mas. The
  differences between the 2MASS and VLBI positions have rms scatter
  $\sigma \approx 114$~mas consistent with the 2MASS position
  accuracy, and the systematic right ascension and declination offsets
  are less than 10~mas.  Of 490 detected sources, 12 have offsets
  exceeding 500~mas (not shown in the plot). Among them, 5 sources are
  considered as good candidates for offset black holes to be
  investigated in depth in follow-up multi-frequency observations.

  The results of anaysis of processed VLBA experiments are accessible from 
the project web page \web{http://astrogeo.org/v2m/}.

\end{document}